\documentclass[reprint,aps,pra,superscriptaddress,showpacs,footinbib,longbibliography,10pt]{revtex4-1}
\usepackage{graphicx} 

\usepackage{color}
\usepackage[colorlinks,bookmarks=false,citecolor=darkblue,linkcolor=red,urlcolor=blue]{hyperref}
\definecolor{darkblue}{rgb}{0,0.02,0.45}

\newcommand{\ilinskite}{KCu$_{5}$O$_2$(SeO$_3$)$_2$Cl$_3$}

\begin{document}

\title{Magnetism of coupled spin tetrahedra in ilinskite-type KCu$_{5}$O$_2$(SeO$_3$)$_2$Cl$_3$}




\author{Danis I. Badrtdinov}
\email{reason2205@yandex.ru}
\affiliation{Theoretical Physics and Applied Mathematics Department, Ural Federal University, 620002 Ekaterinburg, Russia}

\author{Elena S. Kuznetsova}
\affiliation{Department of Chemistry, Moscow State University, 119991 Moscow, Russia}

\author{Valeriy Yu. Verchenko}
\affiliation{Department of Chemistry, Moscow State University, 119991 Moscow, Russia}
\affiliation{National Institute of Chemical Physics and Biophysics, 12618 Tallinn, Estonia}

\author{Peter~S.~Berdonosov}
\author{Valeriy A. Dolgikh}
\affiliation{Department of Chemistry, Moscow State University, 119991 Moscow, Russia}

\author{Vladimir V. Mazurenko}
\affiliation{Theoretical Physics and Applied Mathematics Department, Ural Federal University, 620002 Ekaterinburg, Russia}

\author{Alexander A. Tsirlin}
\email{altsirlin@gmail.com}
\affiliation{Theoretical Physics and Applied Mathematics Department, Ural Federal University, 620002 Ekaterinburg, Russia}
\affiliation{Experimental Physics VI, Center for Electronic Correlations and Magnetism, Institute of Physics, University of Augsburg, 86135 Augsburg, Germany}

\begin{abstract}
Synthesis, thermodynamic properties, and microscopic magnetic model of ilinskite-type KCu$_{5}$O$_2$(SeO$_3$)$_2$Cl$_3$ built by corner-sharing Cu$_4$ tetrahedra are reported, and relevant magnetostructural correlations are discussed. Quasi-one-dimensional magnetic behavior with the short-range order around 50\,K and the absence of long-range order down to at least 2\,K is observed experimentally and explained in terms of weakly coupled spin ladders (tubes) with a complex topology formed upon fragmentation of the tetrahedral network. This fragmentation is rooted in the non-trivial effect of the SeO$_3$ groups that render the Cu--O--Cu superexchange strongly ferromagnetic. 
\end{abstract}

\maketitle

\section{Introduction}
In frustrated magnets, competing spin-spin interactions give rise to unusual types of magnetic order having potential implications for magnetoelectric materials~\cite{mostovoy2007} and complex magnetic textures, such as skyrmions~\cite{okubo2012,leonov2015}. An even more exotic behavior is realized for magnetic ions with spins-$\frac12$ supporting strong quantum fluctuations that keep spins dynamic down to zero temperature and give rise to novel phases of quantum spin liquids~\cite{balents2010,savary2017}. Extensive theoretical research on frustrated spin systems faces a shortage of model compounds that would allow experimental probe of the intricate magnetic phenomena anticipated by theory.

Natural minerals boast highly diverse crystal structures, where different spatial arrangements of the magnetic ions mimic frustrated spin lattices. For example, Cu-based minerals have been instrumental in recent research on the spin-$\frac12$ kagome problem of the two-dimensional (2D) spin lattice of corner-sharing triangles, an enigmatic magnetic model that evades rigorous analytical solution and causes vivid debate regarding the nature of its ground state~\cite{mendels2016,norman2016}. Many other frustrated spin lattices, ranging from simple~\cite{caslin2014} or less than simple~\cite{jeschke2011} spin chains to exotic maple-leaf varieties of the depleted triangular lattice~\cite{fennell2011}, can be realized in the minerals too. 

Cu$_4$ tetrahedra centered by oxygen atoms are a typical building block of copper mineral crystal structures~\cite{krivovichev2013b}. Such tetrahedra can also be viewed as a simple frustrated unit, because they comprise four spin triangles. Here, we report synthesis and magnetic behavior of \ilinskite, a sibling of the mineral ilinskite~\cite{vergasova1997,krivovichev2013}, where Cu$_4$ tetrahedra form layers in the $bc$ plane. Disregarding the tetrahedral picture, the layers can also be viewed as zigzag (sawtooth) chains running along the $b$ direction and bridged by sparse Cu linkers. Given persistent interest in theoretical studies of the sawtooth (delta) chains~\cite{sen1996,nakamura1996,krivnov2014,dmitriev2015} and low-dimensional frameworks of spin tetrahedra~\cite{starykh2002,sindzingre2002,brenig2004,starykh2005,bishop2012,li2015}, as well as the dearth of relevant model materials, we chose to explore magnetic behavior of \ilinskite\ and elucidate its interaction topology. To this end, we combine experimental probes with extensive first-principles calculations, because magnetic interactions in Cu-based minerals are far from trivial~\cite{lebernegg2013,lebernegg2016,janson2016,lebernegg2017}, and \ilinskite\ is no exception.

\section{\label{Results}Results}
\subsection{Synthesis and crystal structure}
Ilinskite is a rare mineral. Its natural samples are too small for most of the experimental probes, whereas previous synthetic attempts reported preparation of only tiny single crystals obtained in a mixture with other copper selenite chlorides~\cite{kovrugin2015}. Therefore, we developed a synthesis method to produce ilinskite-type compounds in larger quantities. Polycrystalline samples of \ilinskite\ were synthesized from binary oxide and chloride precursors in sealed quartz tubes at $380\!-\!400$\,$^{\circ}$C (see Methods for details). X-ray diffraction (XRD) data for such samples are consistent with the crystal structure reported previously~\cite{kovrugin2015}.

An extensive description of the ilinskite-type structures has been given in Refs.~\cite{krivovichev2013,kovrugin2015}. Here, we focus only on those aspects that are germane to the magnetic behavior. In Cu$^{2+}$ compounds, the relevant coordination environment is typically a plaquette formed by four shortest Cu-ligand contacts that define the plane of the magnetic ($d_{x^2-y^2}$) orbital, where $x$ and $y$ are local directions within the plaquette. 

Four crystallographic positions of Cu split into two groups. Cu1 and Cu4 form CuO$_4$-type plaquettes with 4 Cu--O distances of about $1.9-2.0$\,\r A, whereas a distant contact to the Cl atom at 2.59\,\r A (Cu1) and 2.92\,\r A (Cu4) plays no role in the magnetic exchange~\cite{tsirlin2013,jeschke2013,rousochatzakis2015}, because orbitals of the Cl atom do not overlap with the magnetic orbital of Cu$^{2+}$. In the case of Cu2 and Cu3, the plaquettes are of CuClO$_3$ type, which is also not uncommon in Cu-based magnets~\cite{valenti2003,berdonosov2013}. Here, the Cu--O distances are in the same $1.9-2.0$\,\r A range, whereas the Cu--Cl distance is 2.19\,\r A (Cu2) and 2.37\,\r A (Cu3), and $p$-orbitals of the Cl atoms hybridize with the magnetic $d_{x^2-y^2}$ orbital of Cu$^{2+}$.

Viewing the crystal structure of \ilinskite\ from the Cu plaquettes perspective, we find well-defined layers in the $bc$ plane. The layers are bridged by SeO$_3$ groups and additionally interleaved by the K$^{+}$ ions. Each layer can be seen as a sequence of -Cu1-Cu4-Cu2-Cu4-Cu1- zigzag chains along the $b$ direction, with sparse links along the $c$ direction via Cu3. With magnetic interactions restricted to nearest neighbors (Cu--O--Cu bridges), one expects the magnetic topology of spin planes formed by corner-sharing Cu$_4$ tetrahedra (Fig.~\ref{fig:Interactions}). However, Cl atoms are known to mediate long-range superexchange interactions, which render the spin lattice a lot more complex~\cite{valenti2003,tsirlin2009,tsirlin2010}. Our microscopic analysis reported below identifies additional long-range interactions indeed. Even more importantly, dissimilar interactions within the tetrahedra largely relieve the frustration compared to the regular tetrahedral geometry.

\begin{figure}[!h]
\includegraphics[width=0.45\textwidth]{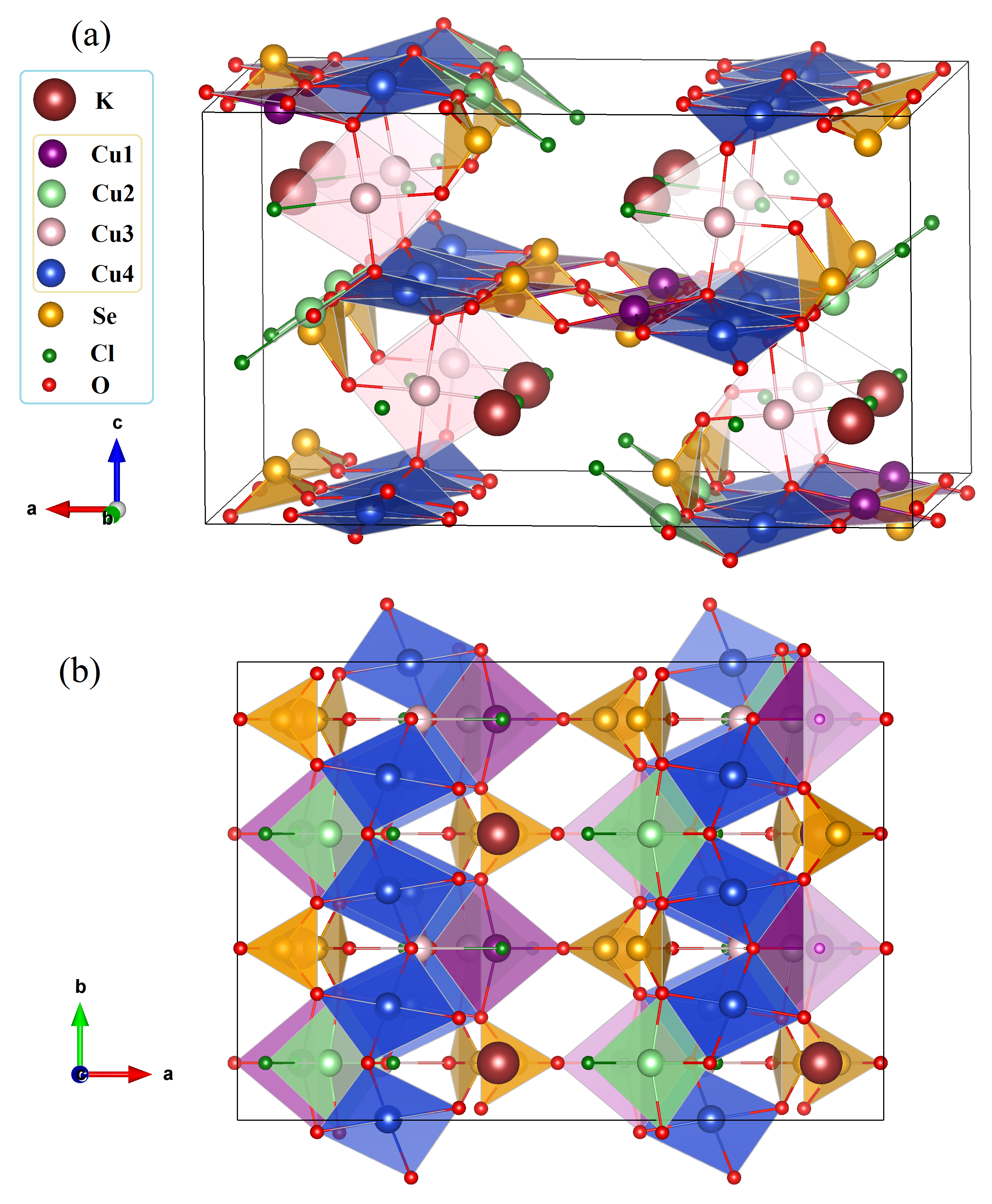}
\caption{(a)-(b): Crystal structure of \ilinskite\ in the $ac$ and $ab$ projections.
 Crystal structures are visualized by using the VESTA software~\cite{VESTA}. }
\label{fig:Crystal}
\end{figure}

\subsection{Thermodynamic properties} 
\begin{figure}[!h]
\includegraphics[width=0.48\textwidth]{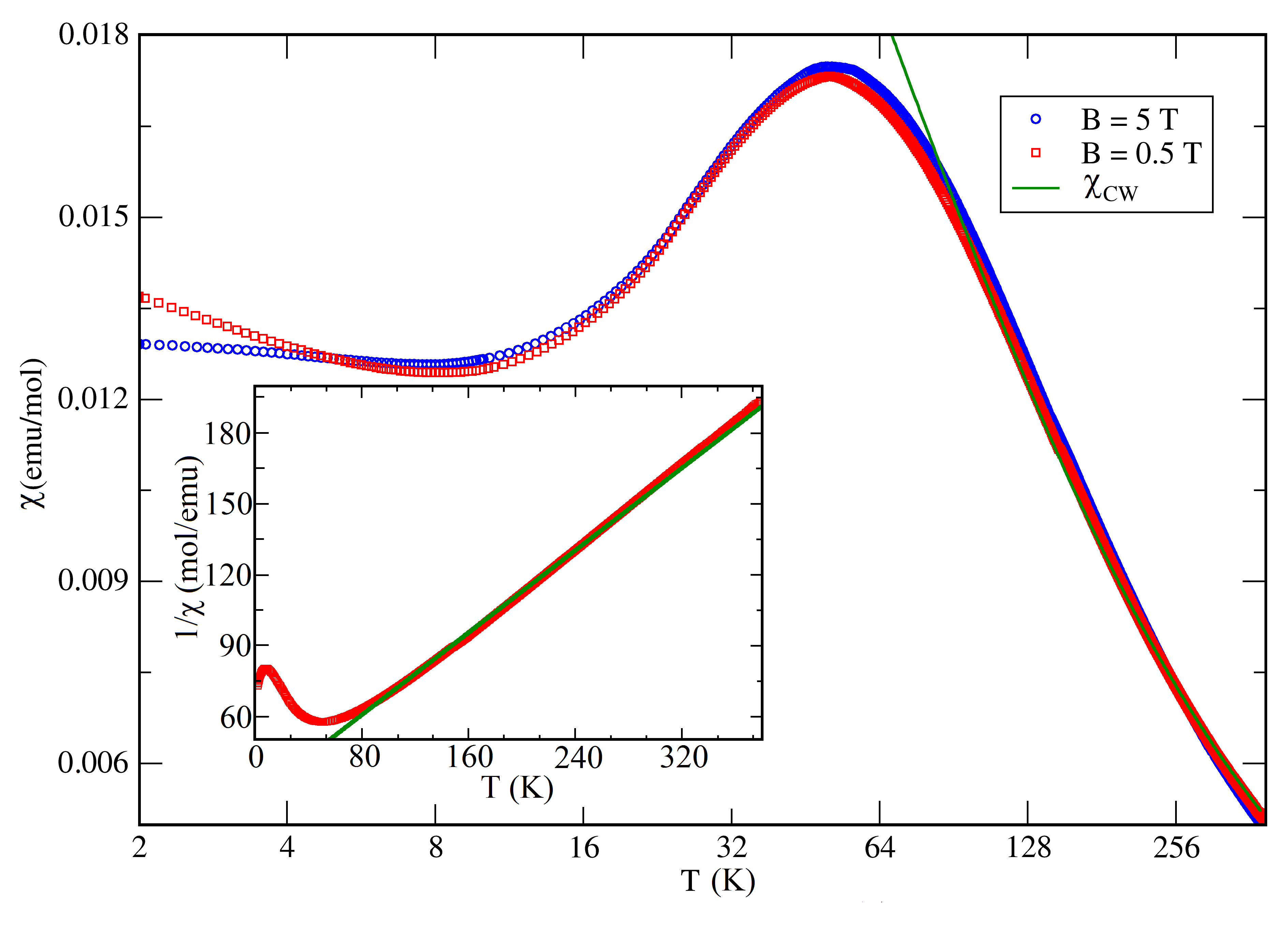}
\caption{Magnetic susceptibility $\chi(T)$ for \ilinskite\ obtained under different values of external magnetic field. The inset shows the Curie-Weiss approximation in 100--380\,K temperature range with parameters $\theta$ = 60\,K and C = 2.3\,emu\,K/mol, as denoted by the green line. }
\label{fig:Susceptibility}
\end{figure}

Magnetic susceptibility of \ilinskite\ shows a broad maximum around 50\,K and a weak upturn below 8\,K (Fig.~\ref{fig:Susceptibility}). The suppression of this upturn in higher magnetic fields indicates its impurity origin. At high temperatures, the susceptibility obeys the Curie-Weiss law $\chi(T)=\frac{C}{T-\Theta}$ with the Curie constant $C=2.3$\,emu\,K/mol and Curie-Weiss temperature $\theta=-60$\,K. The negative value of $\theta$ implies antiferromagnetic (AFM) nature of leading exchange interactions. The $C$ value yields an effective moment of 1.91\,$\mu_B$/Cu, slightly larger than the spin-only moment of 1.73\,$\mu_B$ for Cu$^{2+}$. This leads to an effective $g$-value of $g=2.2$.

\begin{figure}[!h]
\includegraphics[width=0.48\textwidth]{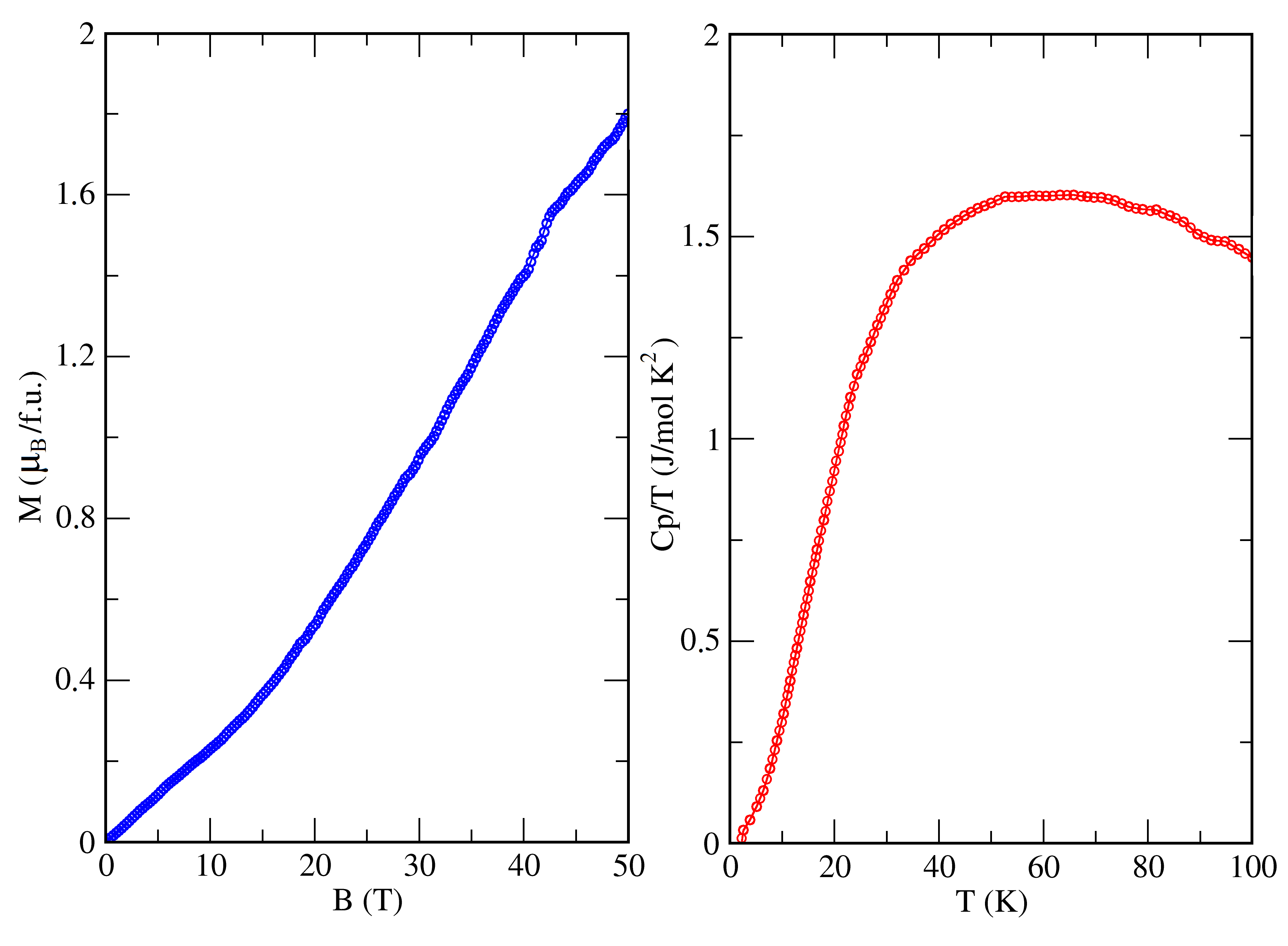}
\caption{(Left panel) The magnetization curve measured at $T=1.5$\,K. (Right panel) Temperature dependence of the specific heat, $C_p(T)/T$, for \ilinskite\ measured in zero field. }
\label{fig:Magnetization}
\end{figure}

The susceptibility maximum around $T_{\max}\simeq 50$\,K indicates AFM short-range order. On the other hand, we do not observe any sharp anomalies that would be indicative of a long-range order setting in at low temperatures. In low-dimensional spin systems, signatures of a magnetic transition are often blurred, because the transition occurs below $T_{\max}$, and the ordered moment is only a fraction of the total magnetic moment~\cite{lancaster2007,tsirlin2012}. Nevertheless, in many of the Cu$^{2+}$ compounds the transitions, even if they occur well below $T_{\max}$, are clearly visible as kinks in $\chi(T)$~\cite{tsirlin2013} or as the divergence of the low-field and high-field susceptibilities~\cite{woodward2002,janson2011b}. This is not the case in \ilinskite, though. Heat-capacity data likewise show no obvious transition anomalies down to 1.8\,K in good agreement with the magnetic susceptibility (Fig.~\ref{fig:Magnetization}, right). 

At the first glance, the susceptibility curve for \ilinskite\ may be reminiscent of a $S=\frac12$ uniform Heisenberg chain (UHC) with the nearest-neighbor antiferromagentic exchange interaction $J$. In such a chain, position and amplitude of the susceptibility maximum yield $\chi^{\rm chain}_{\max}(T_{\max})\,T_{\max}g^{-2}$=0.0353229(3) emu\,K/mol(per Cu) \cite{jonhston2000}. This parameter is independent of $J$, thus providing a simple test whether the UHC model might be applicable. In our case, $\chi^{chain}_{\max}$ at temperature $T_{\max}\simeq 50$\,K is 0.0175\,emu/mol. Using $g=2.2$, we obtain 0.0362\,emu\,K/mol(per Cu) in reasonable agreement with the UHC. However, we show below that the magnetic model of \ilinskite\ is much more involved, and similarities with the susceptibility of the UHC are purely accidental.

Although the susceptibility decreases upon cooling below 50\,K, it does not decay exponentially, as would be expected in a gapped spin system. Magnetization isotherm measured at 1.5\,K reveals a finite slope of $M(H)$ at low fields, which also excludes the presence of a spin gap. The $M(H)$ curve changes slope around 15\,T and shows the increasing trend up to the highest reachable field of 50\,T.

\subsection{\label{sec:Minimal_model_construction}Magnetic model}
For the microscopic description of the magnetic properties of \ilinskite, we construct a minimal Heisenberg-type Hamiltonian that takes into account all the leading exchange interactions between magnetic moments. To this end, we use density functional theory (DFT) methods. 

The structural complexity of \ilinskite\ (Fig.~\ref{fig:Crystal}) is reflected in its intricate electronic spectrum. Indeed, the calculated DFT band structure at the Fermi level obtained with a minimal unit cell is characterized by numerous dispersive and strongly overlapping bands. This band structure is metallic, because it is calculated on the GGA level without taking Coulomb correlations into account. Despite the complexity, one can easily determine the particular copper states producing the bands at the Fermi level. The valence of Cu ions in KCu$_{5}$O$_2$(SeO$_3$)$_2$Cl$_3$ is equal to 2+, placing one unpaired electron to the $d_{x^2-y^2}$ orbital that forms bands in the vicinity of the Fermi level. 

\begin{figure}[!h]
\includegraphics[width=0.48\textwidth]{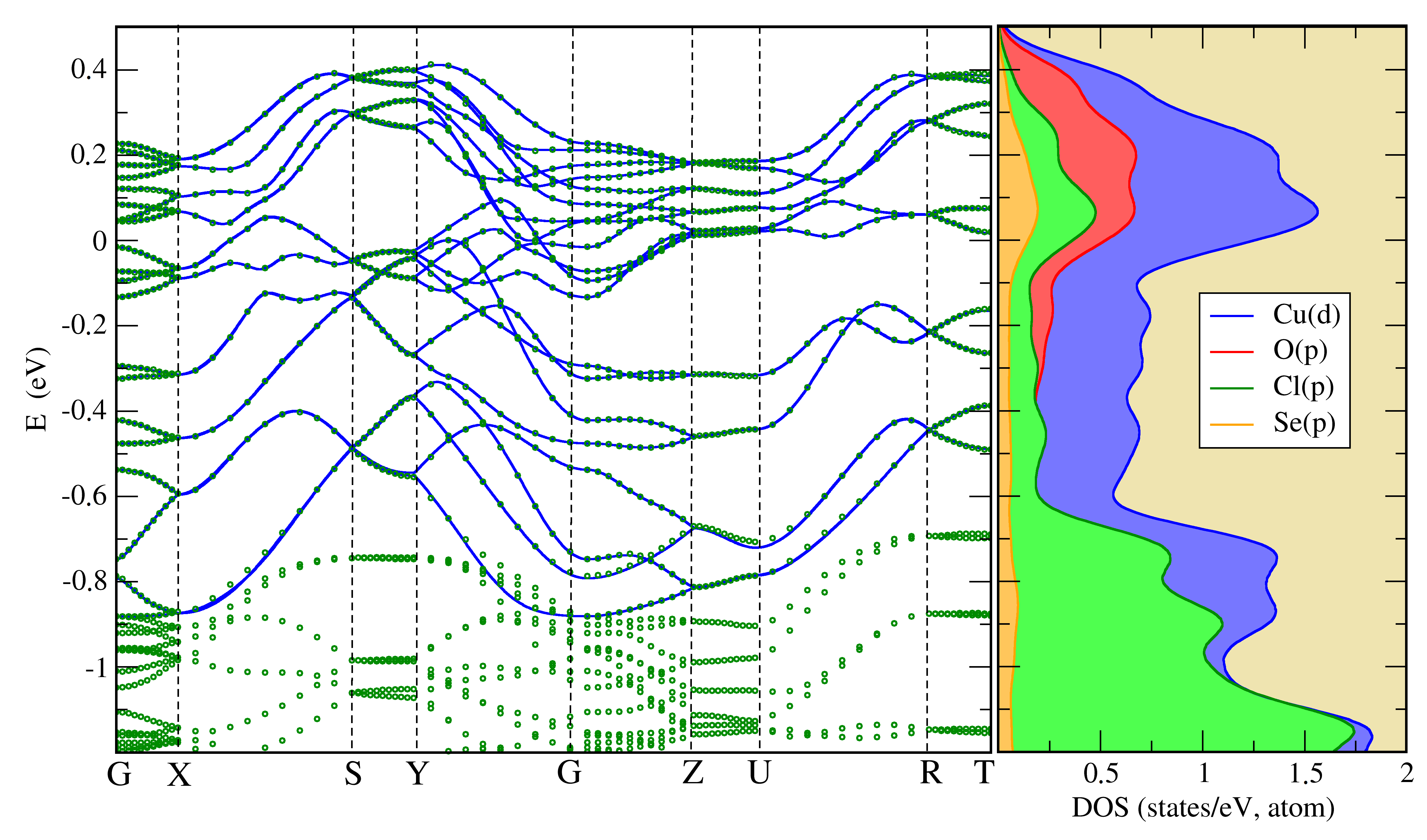}
\caption{(Left panel) Band structure of \ilinskite\ near the Fermi level calculated on the GGA level. The green dotted lines denote the results of the GGA calculation, whereas the blue lines correspond to a minimal tight-binding model constructed in the Wannier function basis. (Right panel) Corresponding atomic-resolved densities of states (DOS).}
\label{fig:Bands}
\end{figure}

We use this fact when constructing the minimal tight-binding model in the Wannier function basis, which gives a preliminary information concerning the magnetic interactions in the system in question. Fig.~\ref{fig:Bands} shows a comparison between the full DFT spectrum and the spectrum of the tight-binding Hamiltonian. The tight-binding model reproduces the DFT solution very accurately. The corresponding hopping integrals between Wannier functions are presented in Table~\ref{tab:hoppings}. Here, we neglect long-range hopping parameters with amplitudes $|t| \le$ 50\,meV. 

Six leading nonequivalent hoppings ($t_{1}$, $t_{2}$, $t_{6}$, $t_{7}$, $t_{8}$ and $t_{11}$) are close to 150\,meV. Five of the underlying superexchange pathways are between nearest neighbors. On the other hand, $t_{11}$ is a long-range interaction between Cu atoms separated by 6.448\,\AA. This clearly identifies the importance of interactions beyond nearest neighbors in \ilinskite. Although AFM contributions to the exchange can be directly expressed as $J_i^{\rm AFM}=4t_i^2/U_{\rm eff}$, with the effective on-site Coulomb repulsion $U_{\rm eff}$, ferromagnetic (FM) contributions are relevant too. Therefore, we restrict ourselves to the 11 potentially relevant interactions listed in Table~\ref{tab:hoppings}, and directly proceed to calculating total exchange couplings $J=J^{\rm AFM}+J^{\rm FM}$ using the DFT+$U$ method, where Coulomb correlations are taken into account on the mean-field level. DFT+$U$ restored the anticipated insulating solution with the energy gap of 4.4\,eV and magnetic moment of 0.75\,$\mu_B$ on copper atoms. 

\begin{table*}[t]
\small
  \caption{Magnetic interactions in \ilinskite: the type of the interacting copper atoms, the Cu--Cu distances $d$ (in\,\r A), the relevant Cu--O--Cu bridging angles (in\,deg), hopping parameters $t_{ij}$ (in\,meV), and total exchange couplings $J_{ij}$ (in\,K) obtained from DFT+$U$. The last column represents the corresponding values, used in QMC simulations. The negative signs of the exchange integrals stand for ferromagnetic interactions. See Fig. \ref{fig:Interactions} for details of the interaction network.}
  \label{tab:hoppings}
  \begin{tabular*}{1\textwidth}{@{\extracolsep{\fill}}c|cccccccc}
 \hline \\
  & $Cu(i) - Cu(j) $  & $d_{\rm Cu-Cu}$ & angle & $t_{ij}$ & $J_{ij}$& $J_{ij}$ & $J_{ij}$ & $J^{\rm QMC}_{ij}$ \\
  &  &         &       &               & $U$=8 eV & $U$=9 eV & $U$=10 eV & \\
\hline \\
1  &   Cu2 - Cu4 & 2.854  & 90,95   & 138.57     &   3.24   &  1.86    &  0.88   &  -       \\ 
2  &   Cu1 - Cu4 & 2.946  & 101, 93 & 158.58     &  14.33   & 10.70    &  7.73   & 6.38     \\
3  &   Cu3 - Cu4 & 3.148  & 113     & $-110.39$  & $-0.45$  & $-1.13$  & $-1.49$ &  -       \\ 
4  &   Cu4 - Cu4 & 3.168  & 114     & $-30.49$   & $-9.52$  & $-7.91$  & $-6.29$ & $-7.91$  \\
5  &   Cu3 - Cu4 & 3.173  & 113     & $-52.33$   & $-10.95$ & $-9.31$  & $-7.68$ & $-5.60$  \\
6  &   Cu1 - Cu3 & 3.174  & 112     & 162.72     &  17.55   &  9.88    &  7.30   & 5.95     \\
7  &   Cu2 - Cu3 & 3.277  & 116     & 160.81     &  13.38   & 10.50    &  8.07   & 6.29     \\
8  &   Cu4 - Cu4 & 3.280  & 121     & $-144.72$  &  10.44   &  7.52    &  5.31   & 7.52     \\
9  &   Cu1 - Cu4 & 6.250  & --      & $-81.37$   &   4.64   &  3.56    &  2.67   &  -       \\ 
10 &   Cu1 - Cu1 & 6.448  & --      & $-96.79$   &   9.48   &  7.62    &  5.98   & 7.62     \\
11 &   Cu2 - Cu2 & 6.448  & --      & $-139.15$  &  17.64   & 14.35    &  11.36  & 14.35    \\
 \hline
  \end{tabular*}
\end{table*}

\begin{figure}[!h]
\includegraphics[width=0.48\textwidth]{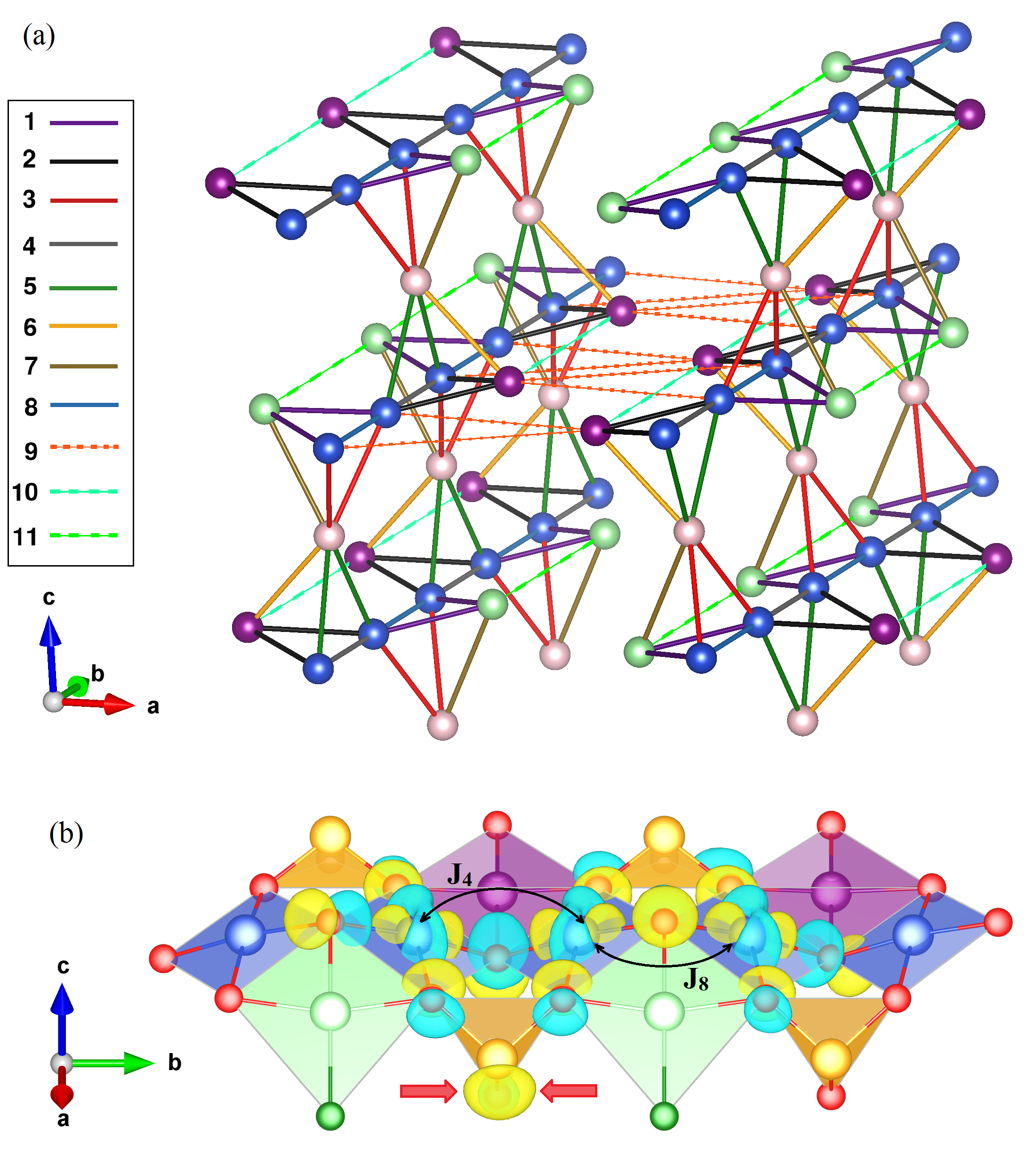}
\caption{(a) Schematic representation of magnetic interactions between the copper atoms in the \ilinskite\ structure. (b) The Wannier function centered on copper atoms within the structural chain. The red arrows indicate the sizable overlap of the Wannier functions at the SeO$_3$ tetrahedra, which leads to difference in the nature of $J_4$ and $J_8$. Different colors denote different phases of the Wannier function.}
\label{fig:Interactions}
\end{figure}

The full set of the isotropic exchange couplings in \ilinskite\ was calculated by a mapping procedure for total energies~\cite{xiang2013,tsirlin2014}. These results are presented in Table~\ref{tab:hoppings}. The change in the $U$ parameter of DFT+$U$ (on-site Coulomb repulsion) leads to a systematic reduction in the magnitudes of $J$'s, because both AFM and FM contributions are reduced when electronic localization is enhanced. The reduction in the AFM part of the exchange is due to the $1/U$ dependence of $J^{\rm AFM}$. The reduction in $J^{\rm FM}$ can be ascribed to the fact that FM superexchange in cuprates depends on the hybridization of the Cu $d_{x^2-y^2}$ orbital with ligand orbitals~\cite{mazurenko2007}, an effect suppressed by the enhanced electron localization at higher $U$'s.

\subsection{Magnetostructural correlations}
The calculated exchange couplings can be divided into AFM ($1-2,6-8$) and FM ($3-5$) sub-groups. Simple magnetostructural correlations rooted in Goodenough-Kanamori-Anderson rules suggest FM superexchange for Cu--O--Cu angles close to $90^{\circ}$ and AFM superexchange away from $90^{\circ}$. This argument explains the $J_2>J_1$ trend, but fails to address peculiarities of other nearest-neighbor couplings that typically feature larger angles but weaker AFM ($J_7,J_8$) or even FM ($J_4,J_5$) exchanges compared to $J_2$. One natural reason for this difference is the presence of two bridging oxygen atoms for $J_1$ and $J_2$ vs. a single oxygen bridge for $J_3-J_8$. However, this does not explain the drastic difference between the strongly AFM $J_6$ with the angle of $112^{\circ}$ and sizable FM $J_4$ and $J_5$ with the even higher angles of $114^{\circ}$ and $113^{\circ}$, respectively.

The twisting of the copper-oxygen plaquettes is another structural parameter relevant to the superexchange~\cite{lebernegg2013}. For example, the superexchange between the orthogonal CuO$_4$ plaquettes can remain FM even if the Cu--O--Cu angle departs from $90^{\circ}$ reaching $100-105^{\circ}$~\cite{nath2013}. One may suggest that this trend persists at even higher bridging angles observed in \ilinskite. However, this twisting argument does not seem to explain peculiarities of our case, because diheral angles between the Cu$^{2+}$ plaquettes for the ferromagnetic couplings $J_4$ ($122^{\circ}$) and $J_5$ ($119^{\circ}$) are larger than that for the antiferromagnetic coupling $J_6$ ($90^{\circ}$). Therefore, the FM couplings occur between less twisted plaquettes, whereas the AFM coupling takes place between the more twisted plaquettes, and, in contrast to Ref.~\onlinecite{nath2013}, the twisting does not enhance ferromagnetism. We thus conclude that side groups should be at play here. Indeed, a closer examination of the crystal structure shows that the FM couplings $J_4$ and $J_5$ are associated with SeO$_3$ links between the copper plaquettes. The coupling $J_6$ lacks such a link and is, therefore, AFM. Likewise, the AFM nature of $J_7$ should be traced back not only to its larger Cu--O--Cu angle compared to that of $J_3-J_6$, but also to the absence of the SeO$_3$ link.  

The effect of the SeO$_3$ groups can be visualized by comparing the interactions $J_4$ and $J_8$ (Fig.~\ref{fig:Interactions},B). The Wannier functions of the copper atoms interacting via $J_4$ have a significant overlap on Se. In contrast, the Wannier functions for $J_8$ do not show such an overlap, and this interaction is restricted to the conventional Cu--O--Cu link. The additional overlap channel may produce the ferromagnetic contribution and eventually lead to the ferromagnetic sign of $J_4$ ~\cite{lebernegg2017}.

Lastly, we discuss the long-range couplings $J_9-J_{11}$. All of them are mediated by the SeO$_3$ groups, as typical for polyanionic compounds, where non-magnetic anions provide shorter O--O distances that are favorable for the Cu--O$\ldots$O--Cu superexchange. Generally the nature of long-range couplings is kinetic, they are strongly dependent on the orbital overlap and, therefore, one the linearlity of the Cu--O$\ldots$O--Cu superexchange pathway measured by the Cu--O--O angle(s). When such a path deviates from linearity, the coupling is suppressed~\cite{janson2011}. In the case of $J_{11}$, the Cu--O--O angle is equal to 170$^\circ$. For $J_{10}$ and $J_{9}$ еру corresponding angles are smaller,  163$^\circ$ and 157$^\circ$, respectively. This trend fully captures the hierarchy of the corresponding exchange couplings.

\subsection{\label{sec:QMS_simulations}Comparison to the experiment}
Within the high-temperature expansion of the magnetic susceptibility, the Curie-Weiss temperature $\theta$ can be expressed through the sum of the exchange couplings $J$ in the following form (for the Cu atom $i$): 
\begin{equation}
\theta_i = -\frac{S(S+1)}{3k_B} \sum_j J_{ij},
\end{equation}
where the summation runs over all pairs of copper atom connected by $J_{ij}$, $k_B$ is Boltzmann constant, and $S = \frac{1}{2}$. Having averaged the Curie-Weiss temperatures calculated for nonequivalent Cu sites within the unit cell, we obtain $\theta$ = $- 91$\,K, $-60$\,K, and $-38$\,K for $U=8$\,eV, 9\,eV, and 10\,eV, respectively. Comparing the theoretical estimates to the experimental value of $\theta=-60$\,K, we find best agreement for the set of $J$'s calculated with $U=9$\,eV.  

\begin{figure}[!h]
\includegraphics[width=0.48\textwidth]{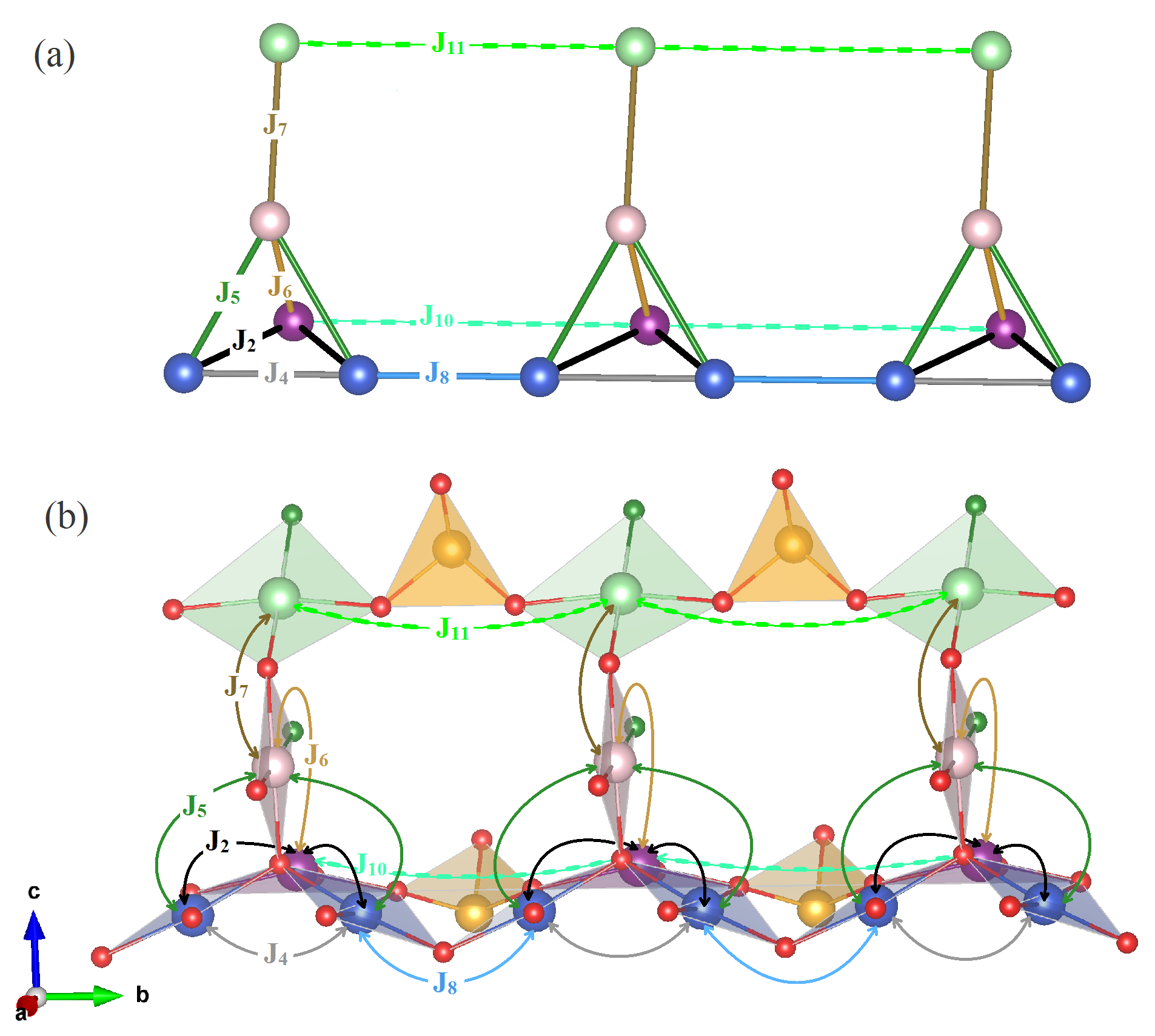}
\caption{(a) Magnetic model used in the QMC simulations. (b) The exchange interactions within the crystal structure.}
\label{fig:Model}
\end{figure}

The spin lattice of \ilinskite\ is frustrated due to the triangles formed by the AFM interactions $J_1-J_1-J_8$ and $J_8-J_9-J_9$. The triangles with two FM couplings $J_3$ and one AFM coupling $J_8$ further contribute to the frustration. This prevents us from simulating magnetic properties of the full three-dimensional DFT-based spin model. However, frustrating interactions are relatively weak comparing to others. Therefore, the non-frustrated model can be introduced as a reasonable approximation when the weaker couplings $J_1$, $J_3$, and $J_9$ are neglected. This decouples the layers of the tetrahedra, because they are connected via $J_9$ only, and further splits the layers into chains with a complex topology (Fig. \ref{fig:Model}). A spin-ladder (tube) motif with $J_2, J_5, J_6$, and $J_7$ acting as legs, and $J_4, J_8, J_{10}$, and $J_{11}$ acting as rungs, can be recognized. The overall geometry is very exotic, though, and clearly lacks any counterpart in theoretical studies of low-dimensional spin systems. Because individual tubes lack magnetic frustration, they are amenable to QMC simulations.

QMC simulations of the magnetic susceptibility reproduce the position of the maximum, but not its amplitude. By varying exchange parameters, we found that the agreement with the experiment can be largely improved if the rung couplings are renormalized by a factor of 0.6. The resulting exchange parameters used in the QMC fit are listed in the last column of Table~\ref{tab:hoppings}. The renormalization can be related to the frustrated nature of \ilinskite. Removing the frustration requires the reduction in at least part of the remaining couplings. 

\begin{figure}[!h]
\includegraphics[width=0.48\textwidth]{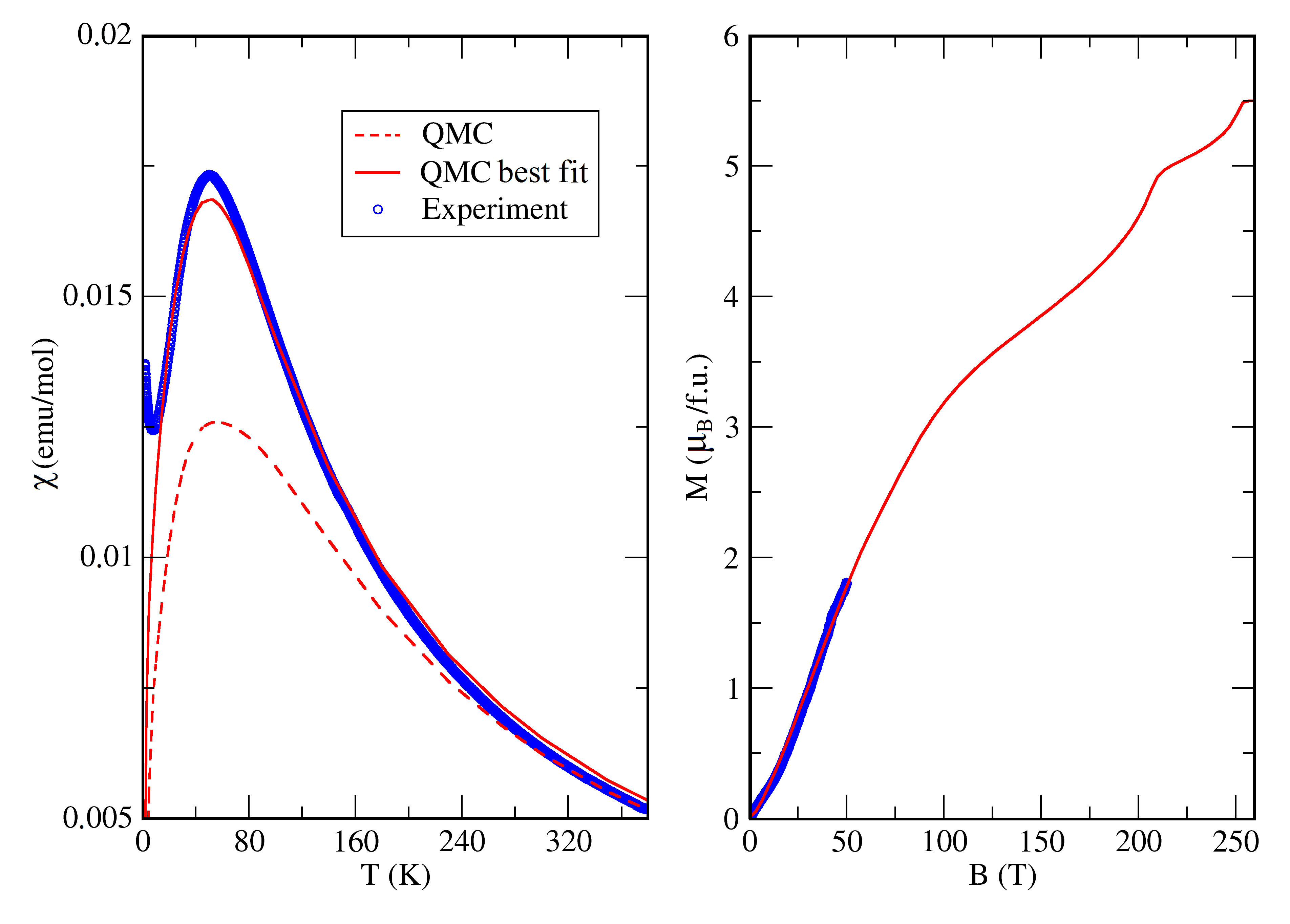}
\caption{(Left panel) The magnetic susceptibility obtained within QMC at 0.5\,T. The straight line corresponds to QMC results with fitted parameters from the last column of the Table~\ref{tab:hoppings}, dashed line -- with parameters directly form DFT results.   (Right panel) The magnetization curve from QMC simulations. }
\label{fig:QMC}
\end{figure}

We also used the exchange couplings from the last column of Table~\ref{tab:hoppings} to simulate the magnetization curve. In agreement with the experiment, we find a steady increase in $M(H)$ up to 50\,T. At higher fields, the curve bends and eventually reaches saturation around 250\,T, the field beyond the reach of present-day pulsed magnets. The 15\,T bend is not reproduced in our simulation. It may be due to anisotropic terms in the spin Hamiltonian, which are beyond the scope of our consideration. It is also worth noting that the simulated curve shows no plateau at zero magnetization, and the gapless nature of the system is well reproduced microscopically. The non-trivial shape of the magnetization curve is likely related to the step-wise saturation of different spins in the lattice. The first bend at $\sim$ 110\,T and around $\frac35$ of the total magnetization is due to the polarization of the three spins connected via ferromagnetic $J_4$ and $J_5$ (Fig.~\ref{fig:Model}). The second bend near $\sim$ 210\,T is likely related to the polarization of the fourth spin in the tetrahedron (suppression of the AFM $J_2$ and $J_6$). Finally, around 250\,T all spins are polarized. 

\section{\label{sec:Summary}Discussion and Summary}
The spin lattice of \ilinskite\ features layers of corner-sharing Cu$_4$ tetrahedra. This relatively simple geometrical motif is amended by the long-range couplings $J_{10}$ and $J_{11}$, but a more drastic effect pertains to the different nature of the bonds on the edges of each tetrahedron. Both FM and AFM exchanges occur between nearest neighbors, and the frustration is largely relieved. Microscopically, this effect originates from a combination of the Cu--O--Cu superexchange pathways and SeO$_3$ bridges that, albeit non-magnetic and seemingly benign, alter Wannier orbitals of Cu$^{2+}$ and affect not only the size but also the sign of the exchange coupling. An equally unanticipated influence of the non-magnetic side groups on the superexchange has been recently reported in the mineral szeniscite~\cite{lebernegg2017}, where MoO$_4$ bridges have an opposite effect to the SeO$_3$ case. They largely enhance an AFM coupling for the bridging angle of about $105^{\circ}$, which is nearly $10^{\circ}$ smaller than the bridging angles for $J_3-J_7$ in \ilinskite.

Extending this analysis to other Cu-based magnets, we realize that the SeO$_3$ groups are often responsible for FM contributions to the exchange. For example, $J^{\rm FM}$ of $-120$\,K was reported in CuSe$_2$O$_5$~\cite{janson2009}, whereas FM interactions between nearest neighbors in francisite, Cu$_3$Bi(SeO$_3)_2$O$_2$Cl~\cite{rousochatzakis2015,prishchenko2017,constable2017}, may also be influenced by the SeO$_3$ links, because the Cu--O--Cu angles are in the same range of $110-115^{\circ}$, where, according to our results, both FM and AFM interactions may occur depending on the presence or absence of the SeO$_3$ link. The SeO$_3$ groups can thus have an indirect, but strong influence on the superexchange, rendering Cu$^{2+}$ selenites an interesting if somewhat unpredictable class of quantum magnets.

\ilinskite\ reveals clear signatures of low-dimensional magnetic behavior. Short-range AFM order is formed around 50\,K, but no signatures of long-range ordering are seen down to 2\,K. This behavior can be rationalized on the microscopic level by the spin lattice comprising robust non-frustrated one-dimensional (1D) units with only weak and frustrated couplings between them. N\'eel temperatures of quasi-1D spin-$\frac12$ antiferromagnets can be orders of magnitude lower than the energy scale of exchange couplings~\cite{janson2009,lancaster2012}. Therefore, it seems plausible that the N\'eel temperature of \ilinskite\ lies below 2\,K. Its detection requires a separate investigation that goes beyond the scope of our present study.

In summary, we prepared single-phase polycrystalline samples of ilinskite-type \ilinskite\ and studied its magnetic behavior. Short-range AFM order sets in below 50\,K, whereas no signatures of long-range magnetic ordering are seen down to at least 2\,K, and no spin gap is observed. This behavior is rationalized microscopically in terms of a non-frustrated 1D spin ladder (tube) with relatively weak and frustrated couplings between the 1D units. The crystal structure of \ilinskite\ features layers of corner-sharing Cu$_4$ tetrahedra. Most of the exchange couplings take place between nearest neighbors, but dissimilar interactions on the edges of these tetrahedra largely reduce the frustration and render the spin lattice quasi-1D. This non-trivial effect originates from an inconspicuous influence of the non-magnetic SeO$_3$ groups that alter superexchange and also mediate long-range couplings.

\section{Methods}

Polycrystalline samples of \ilinskite\ were synthesized using the ampoule technique with KCl, CuO, CuCl$_2$, and SeO$_2$ as reactants. KCl was dried at 140\,$^{\circ}$C prior to synthesis. SeO$_2$ was prepared by dehydration of selenous acid under vacuum ($0.05-0.08$~Torr) and purified by sublimation in the flow of dry air and NO$_2$. Stoichiometric amounts of the reactants were mixed in an Ar-filled glove box. About 1\,g of the mixture was loaded into an evacuated and sealed quartz tube and annealed under the following protocol: i) heating to 300\,$^{\circ}$C for 12 hours; ii) annealing at 300\,$^{\circ}$C for 24 hours; iii) heating to the synthesis temperature $T_{\rm syn}$ for 12 hours; iv) annealing at $T_{\rm syn}$ for 7 days. $T_{\rm syn}$ was varied between 350 and 500\,$^{\circ}$C and had tangible effect on the sample color that varied from emerald green at lower $T_{\rm syn}$ to dark-brown at higher $T_{\rm syn}$. Single-phase samples of \ilinskite\ were obtained at $T_{\rm syn}=380-400$\,$^{\circ}$C and had green color.

Sample quality was checked by x-ray diffraction (XRD) using the STOE STADI-P (CuK$_{\alpha1}$ radiation, transmission mode) and PanAlytical X'PERT III (CuK$_{\alpha}$ radiation, Bragg-Brentano geometry) lab diffractometers. Le Bail fits yield lattice parameters of \ilinskite, $a=18.133(8)$\,\r A, $b=6.438(3)$\,\r A, and $c=10.546(6)$\,\r A. All peaks could be assigned to the ilinskite-type structure, and no impurity phases were found.

Magnetic susceptibility of \ilinskite\ was measured on a powder sample using the vibrating sample magnetometer (VSM) option of the Physical Properties Measurement System (PPMS) from Quantum Design. The data were collected in the temperature range $2-380$\,K under external magnetic fields of $0-14$\,T. Magnetization isotherm up to 50\,T was measured at 1.5\,K in pulsed magnetic fields at the Dresden High Magnetic Field Laboratory. A description of the experimental setup can be found elsewhere~\cite{tsirlin2009b}. The pulsed-field data were scaled using the PPMS data collected below 14\,T.

Magnetic exchange couplings were obtained from first-principles calculations within the framework of density functional theory (DFT) with the generalized gradient approximation (GGA) for the exchange-correlation potential~\cite{pbe96}. To this end, the Quantum Espresso~\cite{espresso} and VASP~\cite{vasp1,vasp2} packages were utilized. The energy cutoff in the plane-wave decomposition was set to 400\,eV, and the energy convergence criteria was chosen at 10$^{-8}$\,eV. For the Brillouin-zone integration a 5$\times$5$\times$5 Monkhorst-Pack mesh was used. The minimal model was constructed in the basis of maximally localized Wannier functions (MLWF)~\cite{Marzari}, where Cu $d_{x^2-y^2}$ states were used as initial projectors.

Exchange parameters $J_{ij}$ of the Heisenberg model
\begin{equation}
\hat{\mathcal{H}} = \sum\limits_{i<j}J_{ij}\hat{\mathbf{S}}_i\hat{\mathbf{S}}_j
\label{eq:Hamiltonian}
\end{equation}
with $S=\frac12$ and the summation over bonds $\langle ij\rangle$, were calculated by a mapping procedure~\cite{xiang2013}. Strong correlation effects were accounted for on the mean-field GGA+$U$ level~\cite{Anisimov} with the on-site Hund's exchange $J_H=1$\,eV and the on-site Coulomb repulsion $U$ varied from 8 to 10\,eV.  

Quantum Monte Carlo simulations were performed using the stochastic series expansion (SSE)~\cite{sandvik1991} method  implemented in the ALPS simulation package~\cite{alps}. Simulations were performed on finite lattice of $N = 1000$ spins $S = \frac{1}{2}$  with periodic boundary conditions. 

\acknowledgements

The work of D.I.B. and V.V.M was supported by the grant of the President of Russian Federation No. MD-6458.2016.2. The work in Tallinn was supported by the European Regional Development Fund, project TK134. AAT acknowledges financial support by the Federal Ministry for Education and Research through the Sofja Kovalevskaya Award of Alexander von Humboldt Foundation. We acknowledge the support of the HLD at HZDR, member of EMFL, and thank Yurii Skourski for the pulsed-field magnetization measurement on \ilinskite.

\end{document}